\documentclass[seceq]{ptptex}
\usepackage{wrapft}

\usepackage{graphicx}

\newcommand{\be}{\begin{equation}}
\newcommand{\ee}{\end{equation}}
\newcommand{\ba}{\begin{eqnarray}}
\newcommand{\ea}{\end{eqnarray}}
\newcommand{\bi}{\begin{itemize}}
\newcommand{\ei}{\end{itemize}}

\newcommand{\nn}{\nonumber \\}

\newcommand{\half}{{\textstyle\frac{1}{2}}}

\newcommand{\<}{\langle}
\renewcommand{\>}{\rangle}
\newcommand{\eq}{Eq.~}

\newcommand{\la}{\label}

\newcommand{\txts}{\textstyle}

\newcommand{\im}{\mathop{\rm Im}}

\newcommand{\Nt}{N_{\tau}}

\notypesetlogo  

\title{
 	Computing the viscosity of the QGP on the lattice
}

\author{
Harvey B. \textsc{Meyer}$^{1,}$\footnote{E-mail: meyerh@mit.edu} }

\inst{
$^1$Center for Theoretical Physics, Massachusetts Institute of Technology\\
77 Massachusetts Avenue, 02139 Cambridge, USA}

\abst{
I review the recent progress made in calculating shear and bulk viscosity
on the lattice, and discuss ways to improve the calculation.
}

\begin{document}

\maketitle

\section{Introduction}

Models treating the system produced in heavy ion collisions at RHIC as an ideal fluid
have had significant success in describing the observed large elliptic flow~\cite{huovinen,shuryak}.
Subsequently the leading corrections due to a finite shear viscosity $\eta$ were 
estimated~\cite{teaney} and led to a remarkably small upper bound on $\eta$.
Recent relativistic viscous hydrodynamics simulations~\cite{Romatschke:2007mq,Song:2007fn,Dusling:2007gi} have started 
to achieve a tighter control over the systematic sources of uncertainty
on the extraction of the shear viscosity to entropy density ratio, $\eta/s$.
On the theory side, 
it is therefore important to compute the QCD shear viscosity from first principles
to complete the picture.
Furthermore, since the heavy ion collision program at LHC will probe the quark-gluon
plasma at temperatures about a factor two higher, it is crucial to predict
the shear viscosity at $\approx3T_c$, and from there to predict the size of elliptic flow,
before experimental data is available. 

A small transport coefficient is a signature of strong interactions;
strong interactions in turn require non-perturbative computational techniques.
In this talk I present a lattice calculation of the
thermal correlators of the energy-momentum tensor
in the Euclidean SU(3) pure gauge theory, and
discuss methods to extract the shear and bulk viscosity from them.
Computationally, the calculation is challenging enough without the inclusion of dynamical quarks,
and physically, the thermodynamic properties of the QGP not too close to $T_c$ 
do not depend sensitively on the flavor content~\cite{karsch-hardprobes06}. 
In perturbation theory~\cite{amy}, the ratio of shear viscosity to
entropy density of the pure gauge theory only differs by about $30\%$
from full QCD~\cite{moore-SEWM04} at a fixed value of $\alpha_s$ 
($\eta/s$ is smaller in the pure gauge theory).
It should be appreciated that this is not much for a quantity which is infinite 
in the Stefan-Boltzmann limit, and the difference is actually reduced when comparing 
the pure gauge theory and full QCD at a common value of $T/T_c$.

Lattice calculations of the shear viscosity~\cite{karsch-visco,nakamura,hmshear}
are based on Kubo formulas (see e.g. \cite{teaney06} and \cite{son-starinets}), 
which relate each transport coefficient to the small frequency behavior
of the retarded correlator. By analytic continuation (see Appendix A), 
the latter is related to the  Euclidean two-point  function of the conserved current.
The significant progress made in~\cite{hmshear}
was to obtain the energy-momentum correlators with an accuracy of about $3\%$
on (isotropic) lattices
with temporal resolution up to $\Nt=12$. This progress was mainly due to the use 
of a multi-level algorithm~\cite{hm-ymills}.

Besides technical improvements that concern the discretization of the correlators, 
I present new ideas to enhance the sensitivity of the lattice data
to the low-frequency region of the spectral function, which determines
the transport properties. Subtracting correlators with
different spatial momentum ${\bf p}$ or different temperature $T$ 
will cancel off most of the contribution of the high-frequency modes. 
The task is then to solve for the difference
of the spectral functions for two different values of ${\bf p}$ or $T$.
This difference is no longer positive definite and solving for the spectral function
requires new methods, such as developed in Ref.~\cite{hmbulk}.

The correlators computed on the lattice are ($L_0=1/T$)
\ba
C_s(x_0,{\bf p})  &=&  L_0^5\int d^3{\bf x} ~e^{i{\bf p\cdot x}} ~\< T_{12}(0) T_{12}(x_0,{\bf x}) \>,
 \nn
C_b(x_0,{\bf p})  &=&  \frac{L_0^5}{9}\sum_{i,j=1}^3
        \int d^3{\bf x}~e^{i{\bf p\cdot x}} ~\< T_{ii}(0) T_{jj}(x_0,{\bf x}) \>.
\la{eq:Cx0}
\ea
For our purposes the spectral functions are then defined by 
\ba
C_{s,b}(x_0,{\bf p}) &=& L_0^5
\int_0^\infty \rho_{s,b}(\omega,{\bf p}) \frac{\cosh \omega(\half L_0-x_0)}{\sinh \frac{\omega L_0}{2}} d\omega.
\la{eq:C=int_rho}
\ea
The shear and bulk viscosities are given by~\cite{karsch-visco}
\be
\eta(T) = \pi \lim_{\omega\to0} \frac{\rho_s(\omega,{\bf 0})}{\omega},\qquad\qquad
\zeta(T) = \pi \lim_{\omega\to0} \frac{\rho_b(\omega,{\bf 0})}{\omega} .
\ee
The spectral functions are positive, $\rho(\omega,{\bf p})/\omega\geq0$,
and odd, $\rho(-\omega,{\bf 0})=-\rho(\omega,{\bf 0})$. If not specified, ${\bf p}$
is set to zero in this talk.
In Ref.~\citen{hmbulk}, I defined the following moments of the spectral function ($n=0,1,\dots$):
\be
 \<\omega^{2n}\>\equiv L_0^5\int_0^\infty d\omega
   \frac{\omega^{2n}\rho(\omega)}{\sinh \omega L_0/2}
       = \left.\frac{d^{2n}C}{dx_0^{2n}}\right|_{x_0=L_0/2}
\la{eq:moments}
\ee
The latter equality implies that they are directly accessible to  lattice calculations.

A first observation is that in infinite spatial volume and in the continuum limit,
\be
C_s(x_0,{\bf p})  = \frac{L_0^5}{4}\int d^3{\bf x}~e^{i{\bf p\cdot x}}
~\< (T_{11}-T_{22})(0) (T_{11}-T_{22})(x_0,{\bf x}) \>,\quad {\bf p}=(0,0,p).
\ee
All the results concerning $C_s$ presented in this talk have been obtained by discretizing this 
form. Secondly, because $ T_{ii} = T_{\mu\mu} - T_{00}$,
and because $\int d^3{\bf x} \<T_{00}(x) {\cal O}(0)\>_c = T^2\partial_T \<{\cal O}\>_T$
for any local operator ${\cal O}$ and $x_0\neq 0$,
\be
\int d^3{\bf x} ~\< T_{\mu\mu}(x)~T_{\nu\nu}(0)\>_c = 
 T^2\partial_T (\epsilon-6P)
+\int d^3{\bf x} ~\< T_{ii}(x)~T_{kk}(0)\>_c\,.
\la{eq:TmumuTii}
\ee
We have used $\< T_{ii} \>_{T-0} = -3P$, $ \<T_{00}\>_{T-0} = \epsilon$. It is therefore convenient
to study the two-point function $C_\theta$ of the trace anomaly $\theta\equiv T_{\mu\mu}$.

\section{Thermal correlators from isotropic lattices}
Figures (1,2) show the ${\bf p=0}$ scalar and tensor correlators for the range of 
temperatures $T_c$ to $3.2T_c$. This data was obtained on $\Nt=8$ lattices with the Wilson 
action and the `bare-plaquette' discretization~\cite{gluex} of the energy-momentum tensor,
and I implemented the tree-level improvement~\cite{hmshear}. The 
leading cutoff effects are thus O($g_0^2a^2$).

The tensor correlator exhibits near-conformal 
behavior, while in the scalar case large departures from conformality are seen, 
particularly near $T_c$. The curves are the leading-order perturbative results. For 
the scalar correlator~\cite{hmbulk}, which is O($\alpha_s^2$),
a choice has to be made for the value of the coupling.
The value  that matches the LO prediction for $\epsilon-3P$ with 
the non-perturbative value~\cite{boyd96}  at $3.22T_c$ is 
$\alpha_s(2\pi T)=\alpha_s^\star\equiv0.289$~\cite{hmbulk}. 
I then use the one-loop evolution on Fig. (2), 
$\alpha_s(2\pi x_0^{-1}) = \alpha_s^\star/(1-\frac{11}{2\pi}\alpha_s^\star \log(Tx_0))$.
The non-perturbative correlator is somewhat flatter than the LO perturbative prediction.
A study of cutoff effects~\cite{hmbulk} for $T_c<T<2T_c$ shows that the $\Nt=8$ data
at $x_0=L_0/2$ is an overestimate, at most by $20\%$, of the continuum correlator.
Large finite-size effects can be excluded on the basis of the large aspect ratio, $LT=6$.


\begin{figure}[htb]
 \parbox{\halftext}{
\centerline{\includegraphics[width=5.4cm,angle=-90]{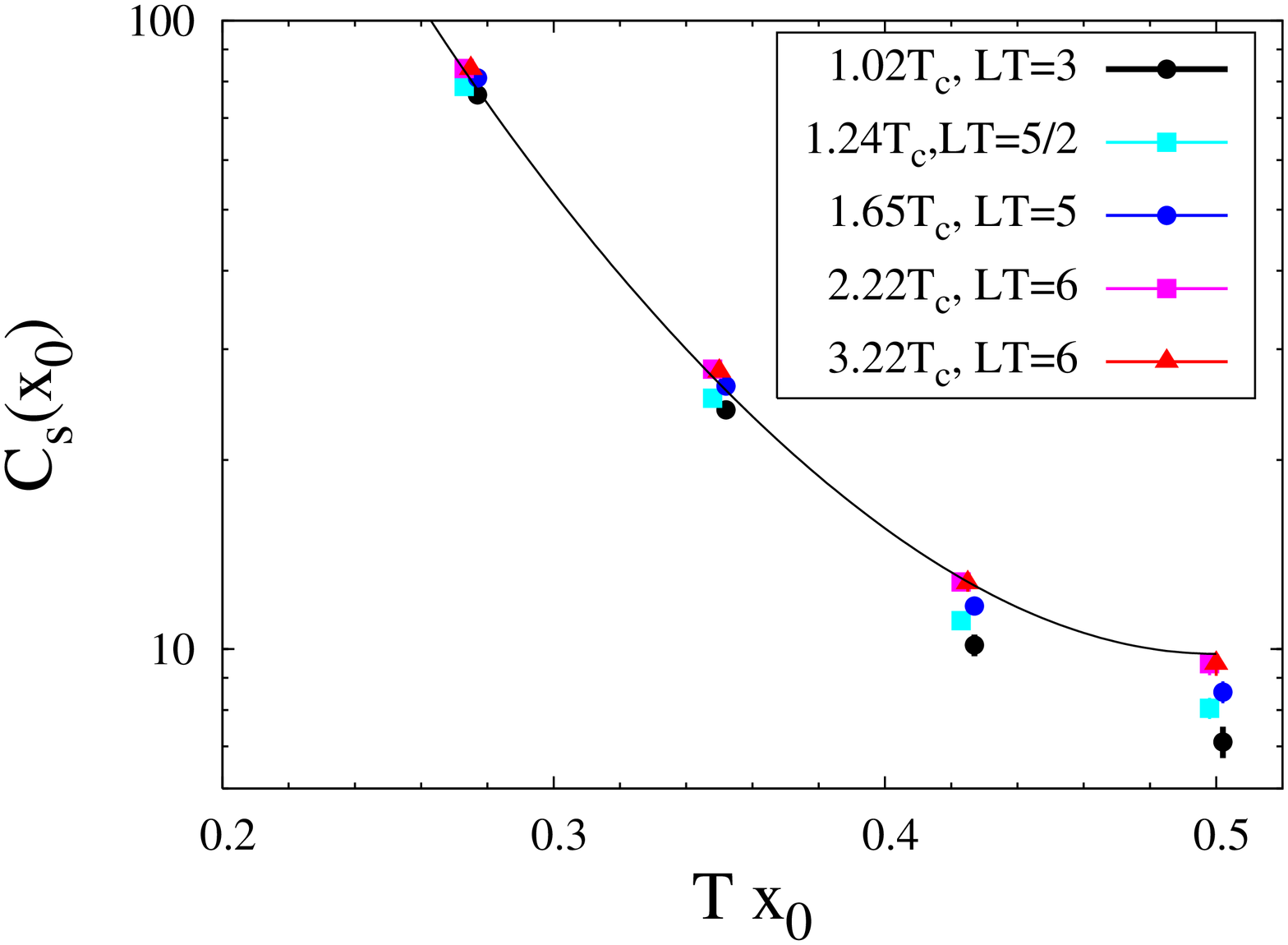}}
                \caption{The tensor correlator $C_s$ for different temperatures}}
 \hfill
 \parbox{\halftext}{
\centerline{\includegraphics[width=5.4cm,angle=-90]{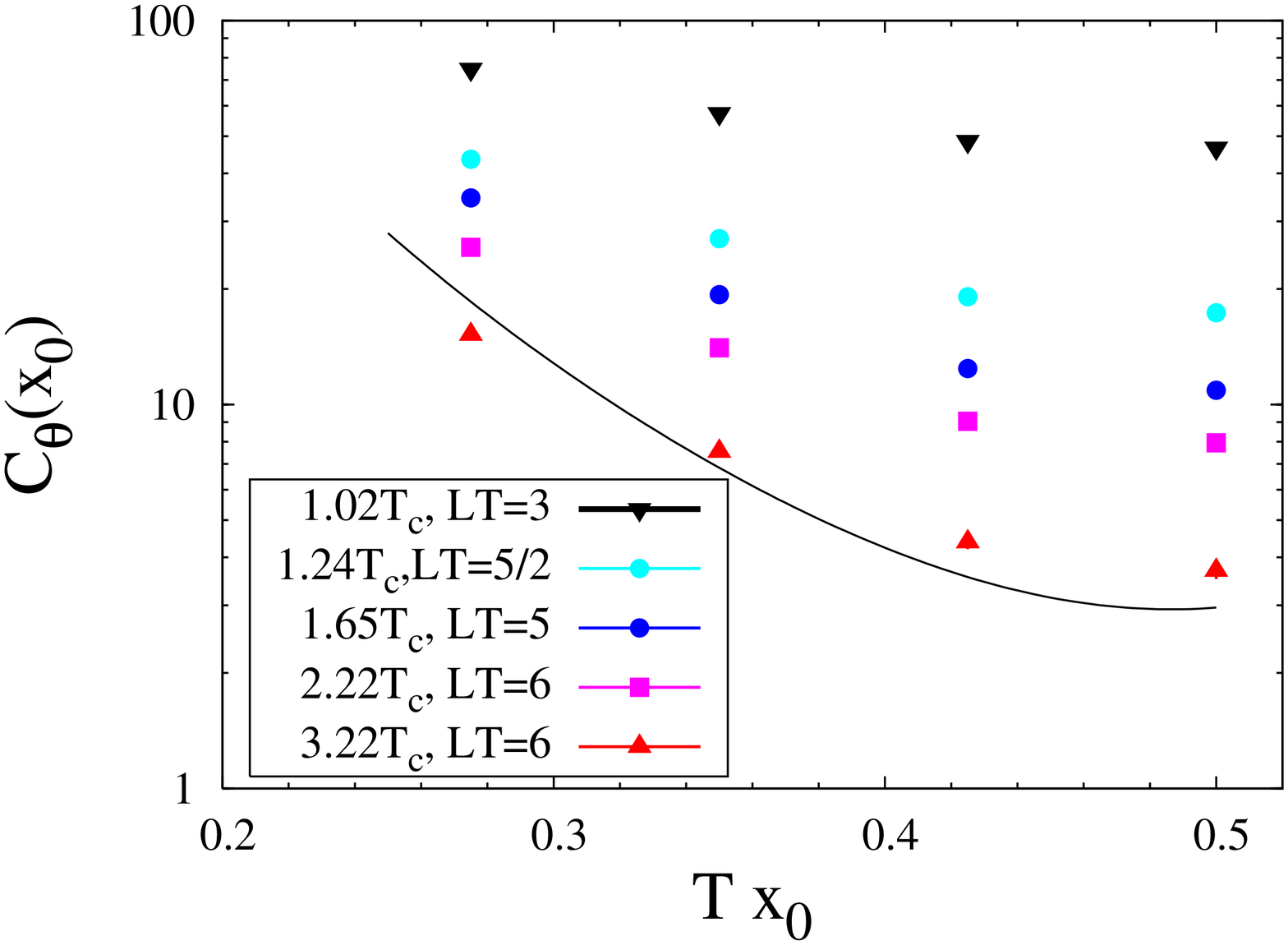}}
                \caption{The scalar correlator $C_\theta$ for different temperatures.}}
\label{fig:tempdepd}
\end{figure}

Figure (3) displays the first two moments, \eq(\ref{eq:moments}), 
of the tensor spectral function at ${\bf p}=0$. They are normalized by the 
leading perturbative result~\cite{hmshear}. I also show the corresponding 
ratio of moments in the (conformal) ${\cal N}=4$ SYM theory~\cite{teaney06,kovtun-starinets}, 
computed by AdS/CFT methods. It is remarkable that 
the typical size and the sign of the deviations from the free approximation is the same
in both theories. 

Based on the data displayed on Fig. (1,2)
and data at smaller spacing, I estimated~\cite{hmshear,hmbulk} the 
shear and bulk viscosity by expanding the spectral function linearly 
in a set of orthogonal functions  $u_\ell(\omega)$.
In this way I obtained for instance $\eta/s = 0.13(3)$ at $1.65T_c$. 
Due to the small size $N_{\rm p}$ of the set, 
the functions fail to satisfy the completeness relation
by an amount quantified by the resolution function 
$\widehat\delta(\omega,\omega')=\sum_{\ell=1}^{N_{\rm p}}u_\ell(\omega) u_\ell(\omega')$.
Figure (4) shows the resolution function for $\omega'/T=0$, 10 and 20 for $N_{\rm p}=4$.
It is broad, and quite far from resembling a delta function at $\omega=\omega'$.
Nevertheless, in cases where the spectral function is smooth, such as the strongly coupled SYM 
theory~\cite{teaney06,kovtun-starinets},  the method works well, see Fig.~(4).

Figure (3) demonstrates that even at $1.6T_c$, the deviation of the first moment, 
to which  the transport peak contributes, from the 
non-interacting approximation, is only about $10\%$. This observation had previously been
made in the finite-temperature perturbation theory framework~\cite{aarts} and in 
the strongly coupled SYM theory~\cite{teaney06}. 
It therefore appears necessary to investigate the 
analytic structure of the spectral function to understand where this lack of 
sensitivity to the low-energy degrees of freedom comes from.

\begin{figure}[htb]
 \parbox{\halftext}{
\centerline{\includegraphics[width=5.4cm,angle=-90]{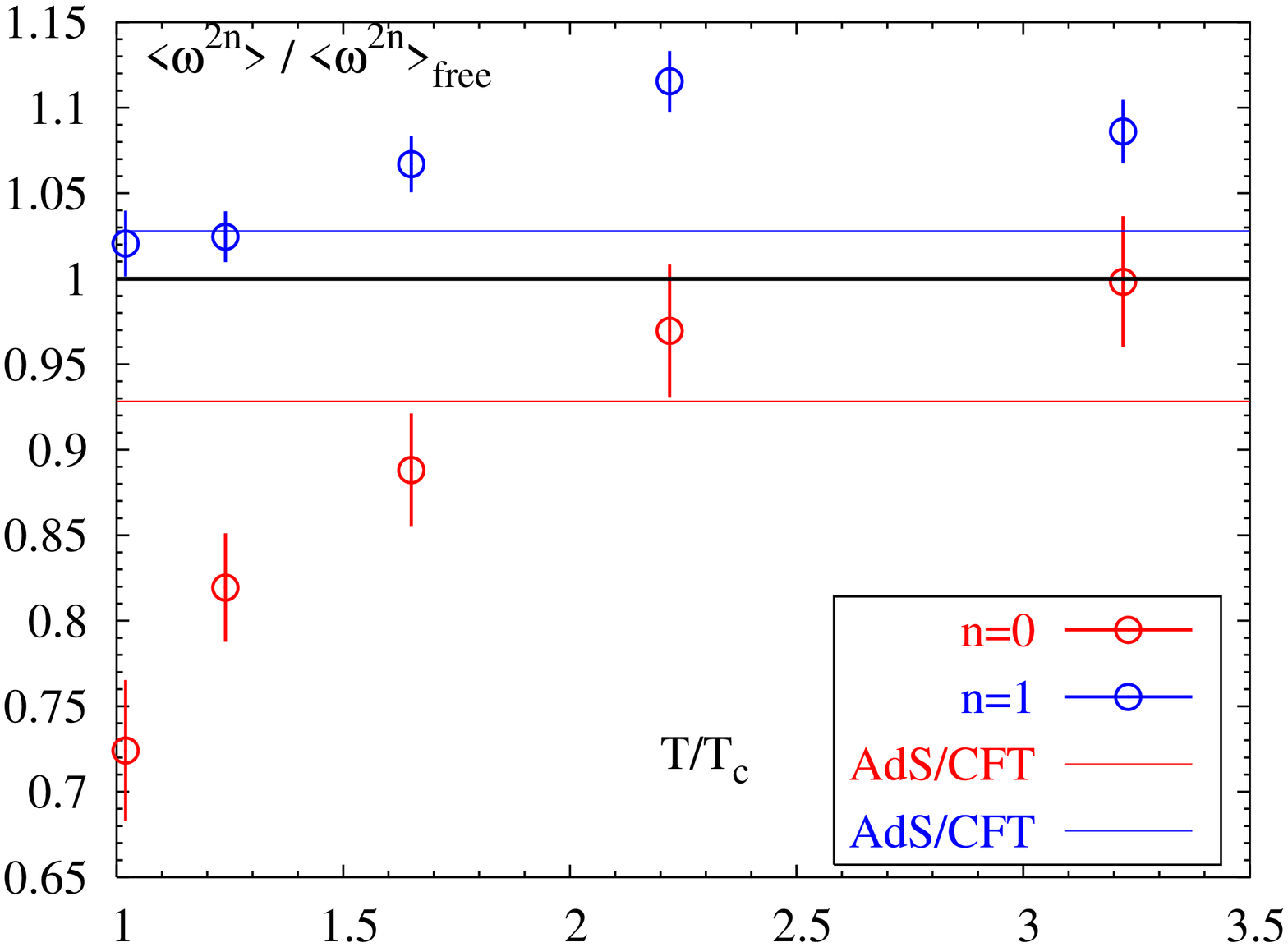}}
  \caption{The lowest two moments of the tensor spectral function 
           for $\Nt=8$ on the isotropic lattice.}}
\label{fig:moments}
 \hfill
 \parbox{\halftext}{
 \centerline{\includegraphics[width=5.4cm,angle=-90]{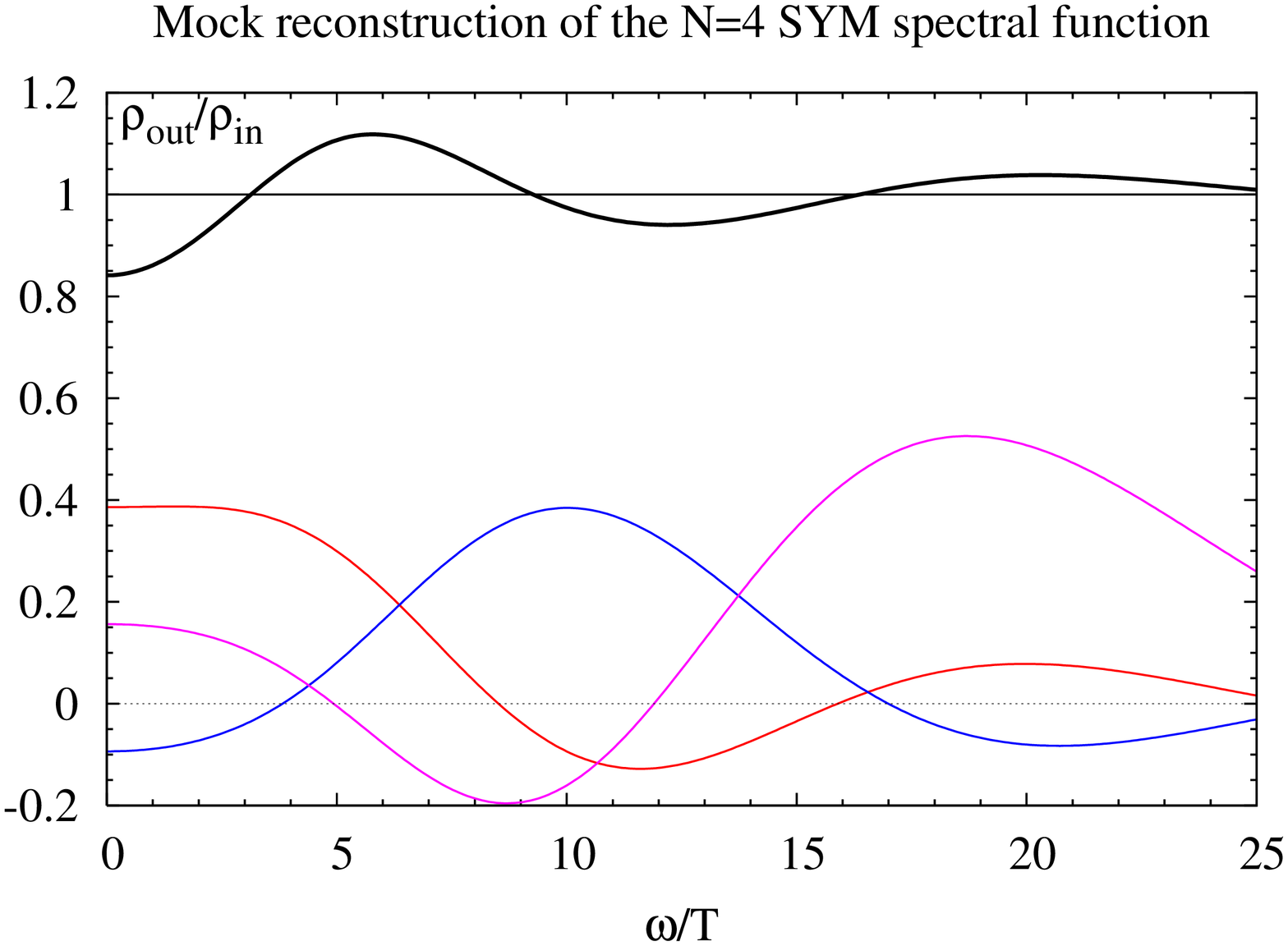}}
  \caption{Reconstruction of the SYM spectral function from 
the $\Nt=8$  Euclidean correlator; resolution functions at 3 points are shown.
}}
\label{fig:N=4}
\end{figure}
\section{Analytic structure of the spectral function}
The leading-order perturbative behavior of the spectral functions can be 
found in Ref.~\citen{hmshear,hmbulk}, they grow as $\omega^4$ at high frequencies,
and at $T=0$ the NLO corrections in $\alpha_s$ are known~\cite{pivovarov,kataev}.
Furthermore, at $T=0$, Lorentz invariance implies
\be
\rho(\omega,{\bf p})= {\rm sgn}(\omega)\theta(\omega^2-{\bf p}^2) ~\rho(\sqrt{\omega^2-{\bf p}^2},{\bf 0}) 
\la{eq:lorentz}
\ee
in the scalar channel.
Based on the operator product expansion~\cite{lissia}, 
we expect the perturbative series of the $\omega^4$ coefficient to be independent of temperature.
Therefore, this term, which makes a large contribution to the Euclidean correlators without
telling us anything about thermal physics, can be eliminated by
subtracting the $T=0$ spectral function from the finite-temperature one.

For low momenta and frequencies, hydrodynamics predicts the functional
form of the spectral functions in the shear channel  and the sound channel
($\rho_{11,11}$, defined as $\rho_s$ with $T_{12}$ replaced by $T_{11}$). For ${\bf p}=(p,0,0)$,
and $v_s$ being the velocity of sound,
\ba
\frac{\rho_{s}(\omega,{\bf p})}{\omega} &= & 
\frac{\eta}{\pi} \frac{\omega^2}{\omega^2+(\eta p^2/(Ts))^2}\,,
\la{eq:12hydro}   \\
\frac{\rho_{11,11}(\omega,{\bf p})}{\omega} & = & \frac{\frac{4}{3}\eta+\zeta}{\pi}
\frac{\omega^4}{(\omega^2-v_s^2p^2)^2+(\omega p^2 (\frac{4}{3}\eta+\zeta)/(Ts))^2}\,.
\la{eq:11hydro}
\ea
See Ref.~\citen{teaney06} for a particularly clear derivation.
It is therefore of interest to study also correlators with non-vanishing
spatial momentum. Ultimately, observing this structure in the spectral function
is the best way to give us confidence in the extraction of the viscosities.

Finally, we remark on a subtlety in the calculation of bulk viscosity.
On general grounds, $\rho_{b}$ 
is expected to \emph{not} have any delta function at $\omega=0$
in an interacting theory.
This would indeed reflect the conservation of (part of) the momentum current,
which would imply in particular that such a current
never dissipates and the system never reaches equilibrium. 
The spectral function for the $C_\theta$ correlator, in view of \eq(\ref{eq:TmumuTii}), must 
then contain the term $T\partial_T(\epsilon-6P)\omega\delta(\omega)$. This singular 
term was missed in~\cite{hmbulk}.

\section{Methods to enhance the sensitivity to the low-frequency region.}
\subsection{Strategy I: exploiting the $T=0$ spectral function}
The idea is to solve the integral equation (\ref{eq:C=int_rho}) for the linear combination
\be
\Delta\rho(T,\omega)\, \equiv \,\rho(T,\omega)-(\rho(0,\omega)-\rho_{\rm 1p}(\omega)).
\ee
 In words, subtract the 
zero-temperature spectral function, except for its one-particle contributions.
Indeed there are two glueballs below the two-particle threshold in both the scalar and the
tensor channel. In infinite spatial volume, 
the function subtracted from $\rho(T,\omega)$ is exactly zero below $2M_{0^{++}}\simeq 3$GeV.
The low-frequency region is therefore unaffected, but the high-frequency 
asymptotics of the function to be reconstructed is now $\omega^0$, up to logarithms.
This is a dramatic improvement.

The first step is thus to determine the $T=0$ spectral function, on which Lorentz symmetry
places much stronger constraints.
The Euclidean correlators deep in the confined phase are shown on Fig.~(5).
They are computed with the isotropic Wilson gauge action, with $a\approx 0.068$ and 
0.051 fm. The correlators fall off rapidly at large distance, 
where a significant signal is obtained only due to the multi-level algorithm.
The correlators have been treelevel improved~\cite{hmshear,hmbulk}.
After this improvement, the two-loop perturbative prediction~\cite{pivovarov}
of the tensor correlator and the one-loop prediction 
for the scalar channel (both with $\alpha_s$ set to 0.25) compare rather well with the data
at short distance. At large distance $x_0>0.5\,$fm, the data is compared to 
the contribution from $\rho_{\rm 1p}$, namely that
of the two stable glueballs present in each channel. I took their masses
from~\cite{thesis} and computed their matrix elements separately~\cite{hmME}.

\subsection{Strategy II: linear combinations of ${\bf p} \neq 0$ spectral functions}
From \eq(\ref{eq:lorentz}), we expect  that for $\omega\gg T$,
the $\omega^4$ contribution cancels up to two-loop order in the linear combination
\be
\rho(\omega,{\bf 0}) \,-\, {\txts\frac{b}{b-1}}\,\rho(\omega,{\txts\frac{{\bf p}}{\sqrt{b}}})
\,+\, {\txts\frac{1}{b-1}}\,\rho(\omega,{\bf p}),\quad b>1.
\la{eq:lincomb}
\ee
In the tensor channel,
this means that the leading large-$\omega$ behaviour of this linear combination is O($\alpha_s^2\omega^4$).
Figure (6) displays  linear combination (\ref{eq:lincomb}) of the correlators $C_s$
for $b=4$ and ${\bf p}=(0,0,\pi T)$.
A cancellation by almost two orders of magnitude takes place and the data is consistent with zero
at all $x_0$. This is a remarkable fact; by contrast $C_s(x_0,{\bf 0})-C_s(x_0,\half {\bf p})$
does not vanish. Obviously such linear combinations deserve further investigation.
The data on Fig.~(6)  was obtained on an anisotropic lattice with 
the clover discretization~\cite{gluex} of the energy-momnentum tensor, with non-perturbative
normalization factors determined using the thermodynamics data in Ref.~\citen{cppacs}.

\begin{figure}
\parbox{\halftext}{\centerline{\includegraphics[width=5.4 cm,angle=-90]{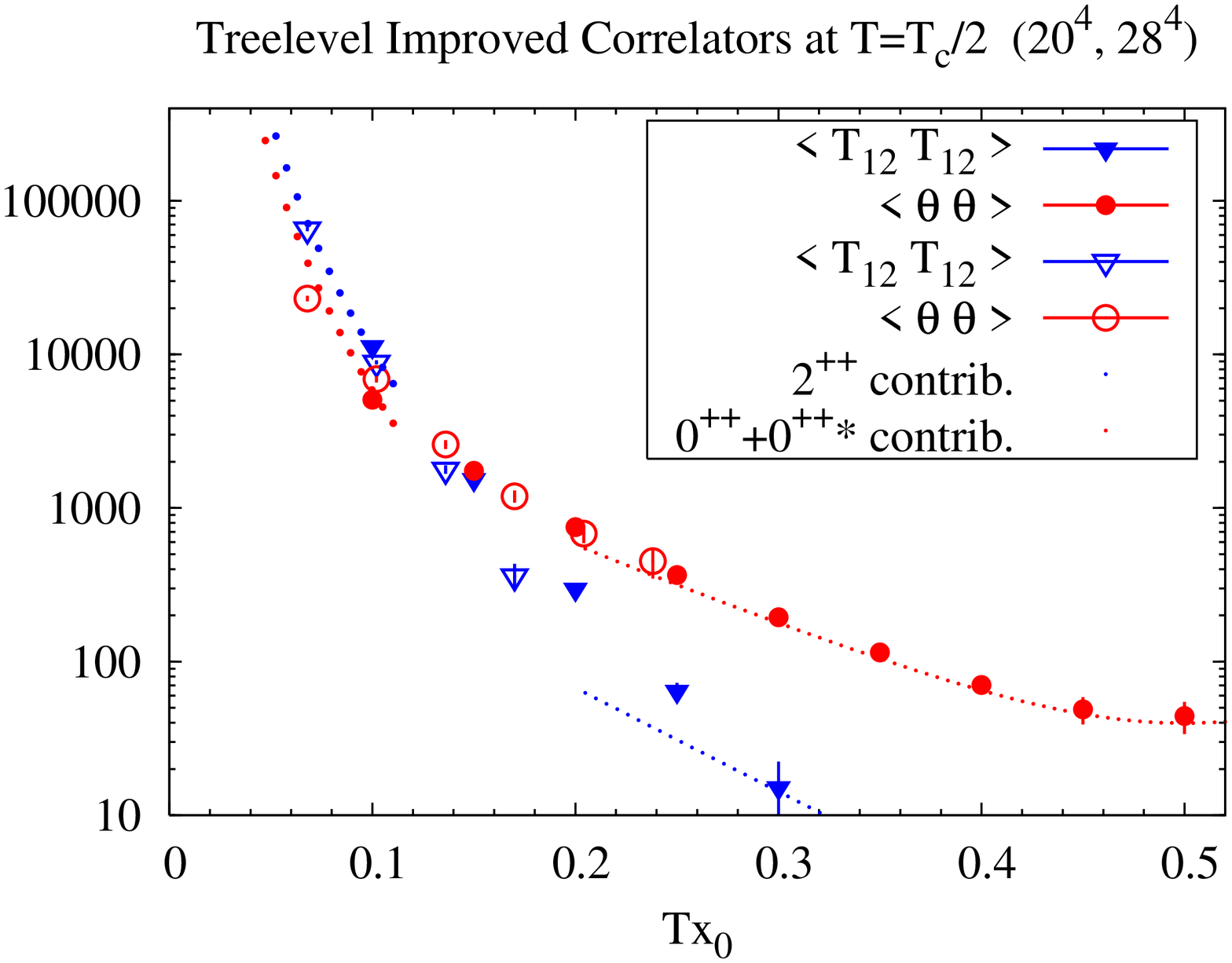}}
\caption{$C_s$ and $C_\theta$ in the confined phase.}}
\hfill
\parbox{\halftext}{\centerline{\includegraphics[width=5.4 cm,angle=-90]{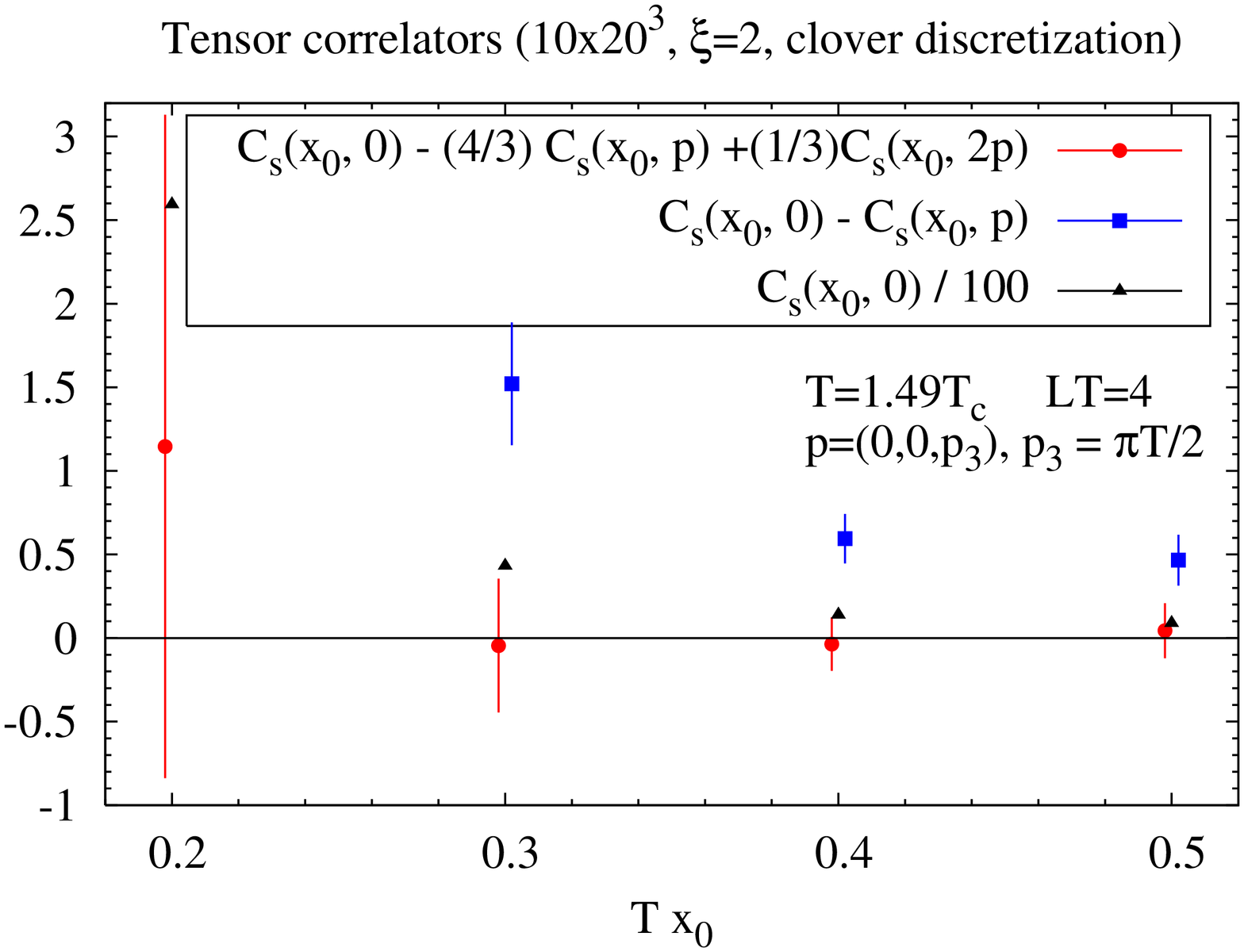}}
\caption{Treelevel improved tensor correlators. }}
\label{fig:T=0;momdiff}
\end{figure}
\noindent

\section{Conclusion}

The  correlators of the energy-momentum tensor 
contain  information  on the plasma that is 
complementary to that of thermodynamics. They are sensitive to
the low-energy degrees of freedom rather than the bulk 
of them, in particular they contain the information on
the transport properties of the system.

Describing the Euclidean data with a smooth spectral function consistent
with positivity, parity and the perturbative large-frequency prediction
leads to a low shear viscosity to entropy ratio~\cite{hmshear}, $1<4\pi \eta/s < 2$
in the temperature range $1.2<T/T_c<1.7$.

In order to increase the sensitivity of the Euclidean correlators
to the low-frequency domain of the spectral function, 
I have proposed two different strategies that aim at subtracting
the contributions of high-frequency modes. The goal is to challenge
the smoothness assumption made on the spectral function~\cite{hmshear,hmbulk}.

A central question is whether perturbation theory can explain
most properties of the plasma at $\approx 3T_c$, a typical temperature probed
at LHC. If it is to account correctly for the viscosities, then it should accurately 
describe the corresponding Euclidean correlators. 
In the scalar channel at $3.2T_c$, the agreement of LO perturbation theory
with the lattice data depends sensitively on the choice of coupling value.
Figure (3) suggest that the agreement is good in the tensor channel at $3.2T_c$, 
however the data is also compatible with a strongly coupled scenario, as the 
qualitative comparison with AdS/CFT reveals. The methods I presented
to subtract the UV-contributions have the potential to elucidate
which is the more appropriate picture of the plasma at temperatures typical
of the LHC experiments.

\section*{Acknowledgements}
\noindent I am grateful to the organizing committee of the New Frontiers in QCD 2008 workshop, 
in particular to Kenji Fukushima, for the kind invitation extended to me. Of all the interesting discussions
I enjoyed at the workshop, I would like to especially mention those 
with G.~Aarts, B.~Mueller, D.~Teaney, R.~Venugopalan that were directly relevant to this work.

\appendix
\section{Analytic continuation of the retarded correlator}
Our goal in this section is to relate the spectral function of a conserved 
operator, defined via a Euclidean correlator, to the imaginary part of
the retarded Green's function in frequency space. A Kubo formula 
relates the latter to a transport coefficient of the finite-temperature system,
in the case of a conserved operator; for the shear viscosity, 
$ \eta = -\lim_{\omega\to0}\frac{1}{\omega} \im G^{12,12}_R(\omega)$,
where $G^{12,12}_R$ is the retarded Green's function of $T_{12}$ 
(see Ref. \citen{teaney06,son-starinets}).

The Euclidean correlator,
$C_E(t)  = \< {\cal O}(t)  {\cal O}(0) \>$, $t>0$,
has the spectral representation
\be
C_E(x_0) = \frac{1}{Z} \sum_{n,m} |{\cal O}_{nm}|^2 e^{-L_0 E_n} e^{E_{nm} x_0}.
\ee
Here ${\cal O}_{nm} = \< n|{\cal O}| m\>$, $E_{nm} = E_n-E_m$.
One easily finds that $C_E(t)$ can be expressed in terms of
the spectral function $\rho(L_0,\omega)$ (see \eq\ref{eq:C=int_rho})
\be
\rho(L_0,\omega)=\frac{2}{Z}\sinh(\omega L_0/2) \sum_{n,m}
\delta(\omega-E_{nm}) e^{-(E_n+E_m)L_0/2}  |{\cal O}_{nm}|^2.
\la{eq:rhospec}
\ee

On the Minkovsky side, the retarded correlator
$iG_R(t) = \theta(t)~ \<[{\cal O}(t),{\cal O}(0)]\>$
has the spectral representation
\be
iG_R(t) = \frac{\theta(t)}{Z}\sum_{n,m}
 |{\cal O}_{nm}|^2 e^{-L_0E_n} ~(e^{iE_{nm}t}-e^{-iE_{nm}t}).
\ee
It is related to the Euclidean correlator by the relation
\be
iG_R(t) = \lim_{\epsilon\to0} (C_E(it+\epsilon) -C_E(-it+\epsilon)), \quad t>0.
\ee
A small positive real part in the argument of $C_E$ guarantees the finiteness of 
the expression. In terms of the spectral function, we obtain
\be
iG_R(t) = \lim_{\epsilon\to0} \int_{-\infty}^\infty d\omega~
\rho(L_0,\omega) e^{-i\omega t} e^{-|\omega|\epsilon}, \quad t>0.
\ee
We have exploited the fact that $\rho$ is odd in $\omega$,
a property manifest in \eq(\ref{eq:rhospec}). The Fourier transform of $G_R$,
$ G_R(\omega) = \int_0^\infty dt~ e^{i\omega t} G_R(t)$,
converges if we give its argument a positive imaginary part:
\be
G_R(\omega+i\delta) = -\int_{-\infty}^\infty d\omega' ~
\frac{\rho(L_0,\omega') e^{-|\omega'|\epsilon}}{\omega'-\omega-i\delta}, 
\quad \omega~{\rm real}.
\ee
In particular,
\be
\im G_R(\omega+i\delta) = \int_{-\infty}^\infty d\omega' ~
(-\pi\rho(L_0,\omega') ~e^{-|\omega'|\epsilon})~ \frac{1}{\pi}\frac{\delta}{(\omega'-\omega)^2+\delta^2}
= -\pi \rho(L_0,\omega),
\ee
where we have recognized one of the standard representations of the delta function, and
let $\epsilon\to0$ in the last step.

\end{document}